\documentclass[pdflatex,sn-mathphys-num]{sn-jnl}

\usepackage{graphicx}
\usepackage{amsmath,amssymb,amsfonts}
\usepackage{amsthm}
\usepackage{xcolor}
\usepackage{manyfoot}
\usepackage{booktabs}
\usepackage{algorithm}
\usepackage{algpseudocode}
\hypersetup{hidelinks}
\begin{document}

\title[When Does Synthetic Pose Data Help?]{When Does Synthetic Pose Data Help? A Real-Anchored Synthetic Augmentation Study for Low-Data Human Pose Estimation}

\author[1]{\fnm{Yukang} \sur{Shen}}\email{yshen4@students.kennesaw.edu}
\author[1]{\fnm{Zhiguo} \sur{Liu}}\email{zliu18@students.kennesaw.edu}
\author[2]{\fnm{Yingshu} \sur{Li}}\email{yili@gsu.edu}
\author*[1]{\fnm{Yan} \sur{Huang}}\email{yhuang24@kennesaw.edu}

\affil[1]{\orgname{Kennesaw State University}, \orgaddress{\city{Marietta}, \state{GA}, \country{USA}}}
\affil[2]{\orgdiv{Department of Computer Science}, \orgname{Georgia State University}, \orgaddress{\city{Atlanta}, \state{GA}, \country{USA}}}

\abstract{\textbf{Purpose:} To determine when controllable synthetic pose data helps a real-domain task in a low-data regime. We build a threshold-calibrated synthetic-first pipeline that combines controllable diffusion generation with semantic and structural curation, and compare five training conditions using a shared 280-image real holdout. Real-plus-raw-synthetic training (1,589 real and 479 synthetic images) achieves pose mAP@0.5 = 0.761 and mAP@0.5:0.95 = 0.411, versus 0.746 and 0.389 for real-only training; raw- and curated-synthetic-only conditions reach 0.449 and 0.430 mAP@0.5, respectively. Synthetic augmentation shows a positive directional trend when anchored by real data, but single-seed, unequal-size comparisons do not establish a causal benefit of curation. We specify the controls required for such a claim and release the pipeline configurations used in this study.}


\maketitle

\section{Introduction}

The performance of modern computer vision systems is often limited by the availability of high-quality labeled data, especially in domains where annotation is expensive, slow, or operationally constrained. Diffusion-based and controllable generation methods make it increasingly practical to synthesize additional training data, but the main challenge is no longer only how to generate plausible individual samples; it is how to assemble synthetic data that remains useful when evaluated on real targets~\cite{Kar_2019_ICCV,Ros_2016_CVPR,Roberts_2021_ICCV,NEURIPS2023_f2957e48,Pozzi2024,singh2024syntheticdataneedbenchmarking,kniesel2025activelearninginspiredcontrolnet,chen2025scalingtumorsegmentationbest,fan2023scalinglawssyntheticimages}.

\begin{figure*}[t]
  \centering
  \includegraphics[width=0.95\textwidth]{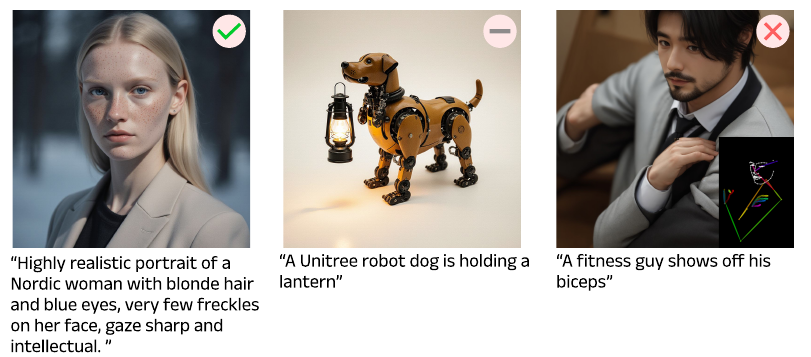}
  \caption{Qualitative comparison under increasing control complexity.
Text-only common prompts are reliable (left), rare prompts degrade semantic fidelity (center), and pose-controlled generation often sacrifices structural coherence (right), illustrating that controllability does not ensure reliability.}
  \label{fig:diffusion-limitations}
\end{figure*}

This distinction matters because controllability at the instance level does not guarantee reliability at the dataset level. Synthetic images may be visually convincing yet still drift from the target distribution, especially for rare poses, compositional prompts, or structurally constrained tasks. As synthetic scale grows, quantity alone can stop helping, making curation and validation as important as generation itself. Fig.~\ref{fig:diffusion-limitations} illustrates this failure mode: text-only prompts are handled reliably, rare compositional prompts degrade in semantic fidelity, and pose-controlled generation can preserve the requested layout while losing structural coherence.

In this work, we present a \emph{Threshold-Calibrated Synthetic-First Data Engine}, a practical framework for expanding data-scarce datasets with controllable generation anchored to a small real dataset. Here, \emph{threshold-calibrated} describes the curation stage: real anchor images provide reference statistics for curation and filtering. This term does not imply a per-sample distribution-matching objective. Our contribution is therefore systems-oriented: we do not propose a new generative model or fine-tuning method, but instead study whether a modular synthetic-first pipeline can make synthetic data useful as low-cost augmentation in real-domain training.

To bridge the reality gap, real images enter the pipeline through threshold calibration: we compute semantic-margin and structural-confidence statistics over a real reference set and set each filter's accept threshold from that distribution. Generated samples are scored and curated before entering the training pool. We do not compute a per-sample distance to the real feature distribution, so the mechanism is threshold calibration rather than distribution alignment; whether an explicit distributional score would select better samples is a question we leave open. The engine also supports optional uncertainty-driven review and human-in-the-loop supervision~\cite{DeepLearningWithAL2022,MosqueiraRey2023HITL}; however, the experiments in this paper evaluate only the generation-and-curation path and do not independently validate those optional feedback components.

This work makes three primary contributions:
(1) We introduce a modular threshold-calibrated synthetic-first data engine integrating controllable generation, two-stage curation, and reproducible data export, released together with the configurations used in this paper.
(2) We present a five-condition pose comparison on a shared real holdout, reporting the direction and magnitude of pose-metric differences under a fixed real-data budget together with the limitations---including unequal total image counts and update counts---that bound their interpretation.
(3) We analyze what this evidence does and does not establish --- including the absence of any measurable filtering benefit at this scale --- and specify the budget-matching and multi-seed controls required before a causal claim about curation is warranted. Our implementation also includes uncertainty-driven review and human-in-the-loop hooks; we do not evaluate them and make no claim about their effect.

The implementation, including the configuration files for every condition reported here, is publicly available at \url{https://github.com/Yan-s-Lab/Data-Engine}.

\section{Related Work}

\subsection{Synthetic Generation Under Real-World Data Scarcity}

Synthetic data has long been used to mitigate data scarcity in computer vision, from graphics-based simulators and domain randomization to modern generative models \cite{tobin2017domainrandomizationtransferringdeep,goodfellow2014generativeadversarialnetworks,song2021denoising}. Diffusion-based models are especially attractive because they offer high visual quality and flexible conditioning, making them practical for low-data augmentation workflows.

To narrow the synthetic--real gap, prior work increasingly anchors generation to real data through real backgrounds, object insertion, or controlled perturbations of real scenes \cite{richter2016playingdatagroundtruth,Ros_2016_CVPR}. Recent surveys review the broader design space of controllable text-to-image diffusion models \cite{Cao_2025}. Controllable diffusion methods further support spatial constraints such as pose, edges, masks, and depth cues \cite{zhang2023controlnet}, while parameter-efficient customization techniques make such generators easier to adapt in practice \cite{hu2022lora,smith2024continualdiffusioncontinualcustomization}. These advances improve sample-level realism and controllability, but they do not by themselves guarantee that a synthetic dataset will be reliable when used at scale.

\subsection{Dataset-Level Reliability and Closed-Loop Curation}

Recent work has therefore shifted attention from individual synthetic samples to dataset-level reliability. Existing studies explore semantic validation, geometric consistency checks, and feature-space metrics for comparing synthetic data with real distributions \cite{SDS,friedman2023knowingdistanceunderstandinggap}. Active learning and human-in-the-loop supervision further help prioritize uncertain or failure-prone samples under limited annotation budgets \cite{kniesel2025activelearninginspiredcontrolnet,DeepLearningWithAL2022,MosqueiraRey2023HITL}.

Our work is aligned with this data-centric view, but focuses on a practical systems question: when a controllable generator, curation module, and downstream learner are connected in a single loop, does the resulting pipeline make synthetic data more useful for real-domain training? The current experiments answer this conservatively: the mixed conditions achieve higher pose mAP than real-only training, while synthetic-only training remains domain-limited and filtering shows no measurable benefit relative to the unfiltered pool at the present scale.

\section{Methods}

\subsection{System Overview}

\begin{figure*}[t]
  \centering
  \includegraphics[width=0.95\textwidth]{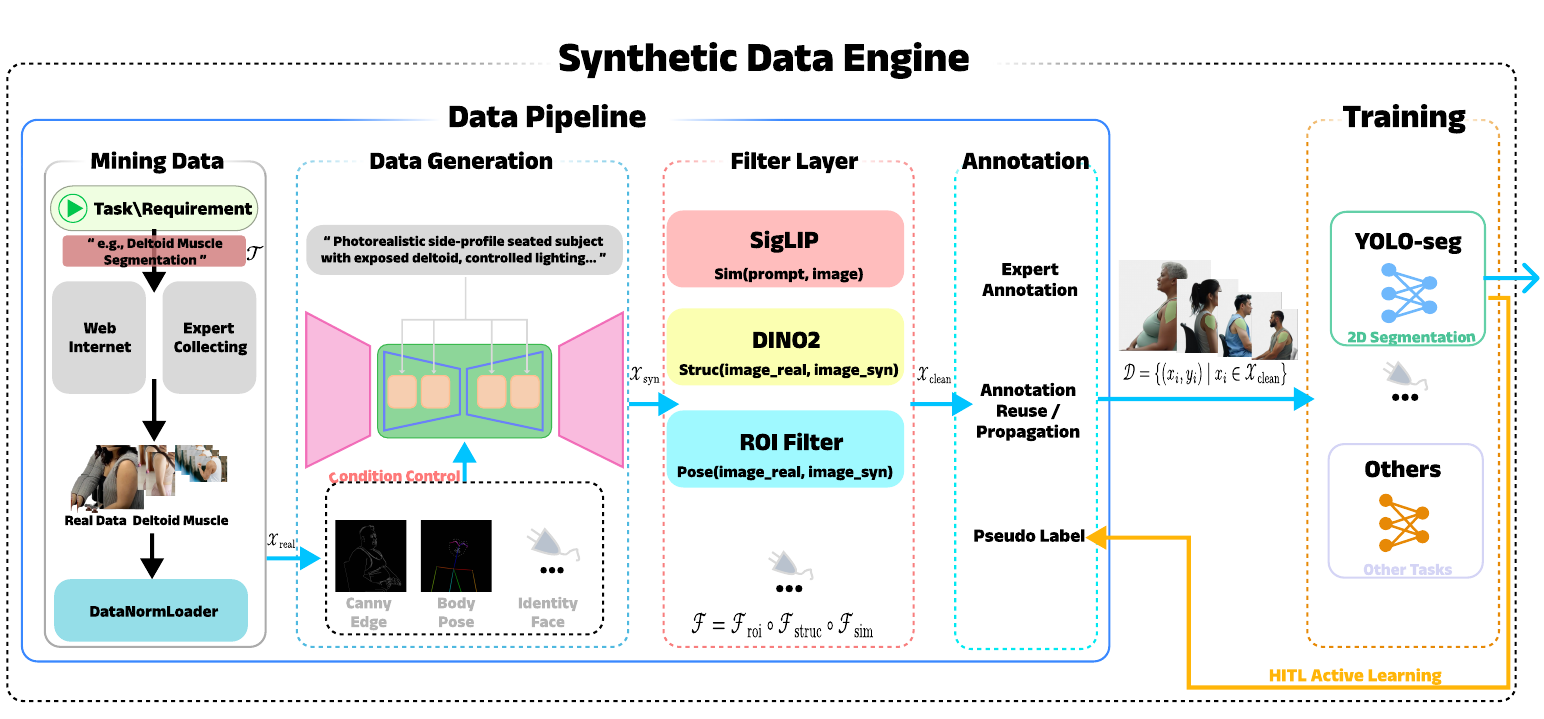}
  \caption{Overall architecture of the proposed Threshold-Calibrated Synthetic-First Data Engine. The system integrates diffusion-based generation and threshold-calibrated multi-stage filtering, while also supporting optional uncertainty-driven selection and human review modules.}
  \label{fig:overall-framework}
\end{figure*}

The proposed Threshold-Calibrated Synthetic-First Data Engine is a task-driven data pipeline rather than a standalone generative model. As illustrated in Fig.~\ref{fig:overall-framework}, it consists of four practical stages: (1) collection of a small real anchor set, (2) controllable synthetic generation using latent diffusion models (LDMs)~\cite{rombach2022highresolution,zhang2023controlnet,smith2024continualdiffusioncontinualcustomization}, (3) threshold-calibrated curation/filtering using semantic and structural acceptance thresholds calibrated on real reference images, and (4) downstream export for model training. Optional annotation refinement and human review are supported by the engine design, but are not part of the core experimental validation in this paper.

\paragraph{Task Initialization and Real Anchoring.}
Given a target task $\mathcal{T}$, a small set of real samples $\mathcal{X}_{\text{real}}$ is collected as domain anchors. These anchors provide reference statistics and structural priors for downstream filtering; when image-based controls are used, they can also provide structured guidance for generation.

\paragraph{Synthetic-First Generation and Curation.}
A controllable diffusion-based generator produces an initial synthetic pool
\[
\mathcal{X}_{\text{syn}} = \{ x_{i,v} \mid i \in \{1,\dots,N\},\; v \in \{1,\dots,K\} \},
\]
where $i$ indexes scene instances and $v$ denotes controlled variation factors such as pose, edge condition, lighting, or viewpoint. The curated synthetic export used downstream is then defined by
\[
\mathcal{X}_{\text{clean}} = \mathcal{F}(\mathcal{X}_{\text{syn}}),
\]
where $\mathcal{F}$ denotes the curation/filtering pipeline. Operationally, $\mathcal{X}_{\text{clean}}$ is the synthetic export retained for downstream training after curation and task-specific post-processing.

\paragraph{Downstream Use.}
The curated synthetic export is then combined with real data for downstream training. In the current paper, this downstream validation is performed through a single-pass pose comparison on a shared real holdout set. Although the engine is compatible with iterative refinement and selective human review, those feedback components are treated here as optional system extensions rather than experimentally verified claims.

Algorithm~\ref{alg:engine} summarizes the main data flow of the proposed pipeline.

\begin{algorithm}[t]
\caption{Threshold-Calibrated Synthetic-First Data Engine}
\label{alg:engine}
\begin{algorithmic}[1]
\Require Real anchor set $\mathcal{X}_{\text{real}}$, task $\mathcal{T}$, generator $p_\theta$, thresholds $\tau_{\text{sem}}, \tau_{\text{struct}}$
\Ensure Curated export $\mathcal{X}_{\text{clean}}$
\State Compute anchor embeddings $\{E(r)\}_{r \in \mathcal{X}_{\text{real}}}$
\State Calibrate or set task-specific thresholds $\tau_{\text{sem}}, \tau_{\text{struct}}$ on labeled real anchors
\State Sample structured control signals $\{c_{\text{prompt}}, c_{\text{pose}}, c_{\text{edge}}\}$ from task-valid distribution
\State Generate $\mathcal{X}_{\text{syn}} = \{x_{i,v}\}$ from $p_\theta(x \mid c_{\text{prompt}}, c_{\text{pose}}, c_{\text{edge}})$
\State $\mathcal{X}_1 \leftarrow \emptyset$
\For{each $x \in \mathcal{X}_{\text{syn}}$}
  \State $s_{\text{sem}}(x) \leftarrow \text{score}_{\text{pos}}(E(x)) - \text{score}_{\text{neg}}(E(x))$
  \If{$s_{\text{sem}}(x) \geq \tau_{\text{sem}}$}
    \State $\mathcal{X}_1 \leftarrow \mathcal{X}_1 \cup \{x\}$
  \EndIf
\EndFor
\State $\mathcal{X}_{\text{clean}} \leftarrow \emptyset$
\For{each $x \in \mathcal{X}_1$}
  \State $s_{\text{struct}}(x) \leftarrow \text{Struct}_{\mathcal{T}}(x)$
  \If{$s_{\text{struct}}(x) \geq \tau_{\text{struct}}$}
    \State $\mathcal{X}_{\text{clean}} \leftarrow \mathcal{X}_{\text{clean}} \cup \{x\}$
  \EndIf
\EndFor
\State \textit{[Optional]} Route borderline or rejected samples to HITL review
\State \Return $\mathcal{X}_{\text{clean}}$
\end{algorithmic}
\end{algorithm}

\subsection{Controllable Synthetic Generation}

Synthetic samples are generated with a pretrained diffusion model used in a training-free manner. Rather than learning a task-specific generator, we treat the model as a controllable sampler driven by structured prompt, pose, and edge conditions:
\[
x \sim p_\theta\!\left(x \mid c_{\text{prompt}}, c_{\text{pose}}, c_{\text{edge}}\right).
\]
Prompt templates and control signals are defined by task-valid constraints, and their combinations are sampled to increase structural coverage rather than to imitate individual real images. Real anchors calibrate control settings and filtering thresholds and may provide image-based structural guidance; we do not fine-tune the generator.

This design keeps the generative component simple and reproducible. The paper therefore does not claim a new generation algorithm; the contribution lies in how controllable generation is embedded into a data-engineering workflow for downstream augmentation.

\subsection{Task-Driven Curation and Filtering}
\label{sec:filtering}

To preserve structural validity and semantic coherence without collapsing diversity, we adopt a two-stage curation cascade. Let $\mathcal{X}_{\text{syn}}$ denote the synthetic pool and $\mathcal{X}_{\text{real}}$ the real anchor set. For each $x \in \mathcal{X}_{\text{syn}}$, we apply
\[
\mathcal{F}_{\mathcal{T}} = \mathcal{F}_{\text{struct}} \circ \mathcal{F}_{\text{sem}}.
\]
The cascade acts as a high-recall verifier rather than a hard capability gate.

\textbf{Filter 1 --- Semantic alignment} ($\mathcal{F}_{\text{sem}}$) is a vision-language margin score computed against task-positive and task-negative text templates calibrated on the real anchor set:
\[
s_{\text{sem}}(x) = \text{score}_{\text{pos}}(E(x)) - \text{score}_{\text{neg}}(E(x)),
\]
where $E(\cdot)$ is a pretrained vision-language encoder (SigLIP2 in our implementation~\cite{tschannen2025siglip2}) and $\text{score}_{\text{pos}/\text{neg}}$ are top-$k$-mean logits over positive and negative prompt sets. Samples below a margin threshold $\tau_{\text{sem}}$ calibrated on labeled real data are rejected.

\textbf{Filter 2 --- Structural validity} ($\mathcal{F}_{\text{struct}}$) combines bounding-box coverage and keypoint confidence to gate samples on task-relevant structure:
\[
s_{\text{struct}}(x) = \text{Struct}_{\mathcal{T}}(x),
\]
implemented as a YOLO pose/ROI gate that checks person bounding-box area ratio and the count of high-confidence keypoints against thresholds $\tau_{\text{struct}}$. Only samples failing clearly on either stage are rejected; borderline cases may be routed to downstream review.

This two-stage cascade operationalizes what we mean by \emph{threshold-calibrated}: real anchors supply the reference statistics from which both filter thresholds are set, rather than a per-sample similarity score against the real distribution. At the same time, the pose experiments in Section~\ref{sec:exp-pose-ablation} show that this component does not yet yield a decisive advantage over raw synthetic data at the current scale.

\subsection{Optional HITL and Feedback Extensions}

The Label Studio HITL loop, uncertainty-based routing, and closed-loop refinement are fully implemented in the codebase (\texttt{pipelines/closed\_loop\_round.py}, \texttt{label/label\_studio\_push.py}) but are not ablated in the current paper. In a full closed-loop deployment, high-uncertainty samples would be routed to annotators or recycled into later data-selection rounds while the generator remains frozen. We present these as extensible system features rather than as experimentally validated contributors to the results in Section~\ref{sec:experiments}.

\subsection{CLI-First Modular Engine Design}
Each stage of the pipeline is implemented as an isolated CLI module with explicit input/output artifacts, building on the ecosystem of tools such as FiftyOne, Label Studio, and Snorkel \cite{voxel51fiftyone,heartexlabelstudio,ratner2017snorkel} for curation, annotation, and labeling primitives. Rather than proposing a new annotation platform, the engine emphasizes orchestration: generation, filtering, evaluation, and optional human review are exposed as auditable, independently swappable CLI stages. This design ensures reproducibility and clear separation of concerns across the pipeline.

\section{Experiments}
\label{sec:experiments}

\subsection{Goal and Evaluation Protocol}
\label{sec:exp-goal}

Our primary evaluation focuses on the \textbf{human pose estimation} task, which serves as the core comparison benchmark of the paper. The goal is not to claim a new pose architecture, but to test whether our threshold-calibrated synthetic-first pipeline can provide \emph{usable augmentation without additional manual keypoint annotation}: synthetic data should be insufficient on its own, yet still improve pose metrics when mixed with real anchors under a controlled training recipe.

All pose conditions are evaluated on a shared \textbf{real holdout set} of 280 images using identical YOLO11s-pose hyperparameters. We report pose $\mathrm{mAP}@0.5$, pose $\mathrm{mAP}@0.5{:}0.95$, box $\mathrm{mAP}@0.5$, pose precision, and pose recall. The shared holdout keeps the evaluation target fixed across conditions, but it does not isolate a causal effect of synthetic data because data composition is confounded with total image count and optimizer updates. Accordingly, our central claim is limited to a directional pose-metric comparison: under this protocol, the mixed conditions achieve higher pose mAP than real-only training.

\begin{figure*}[!htbp]
  \centering
  \includegraphics[width=0.90\textwidth]{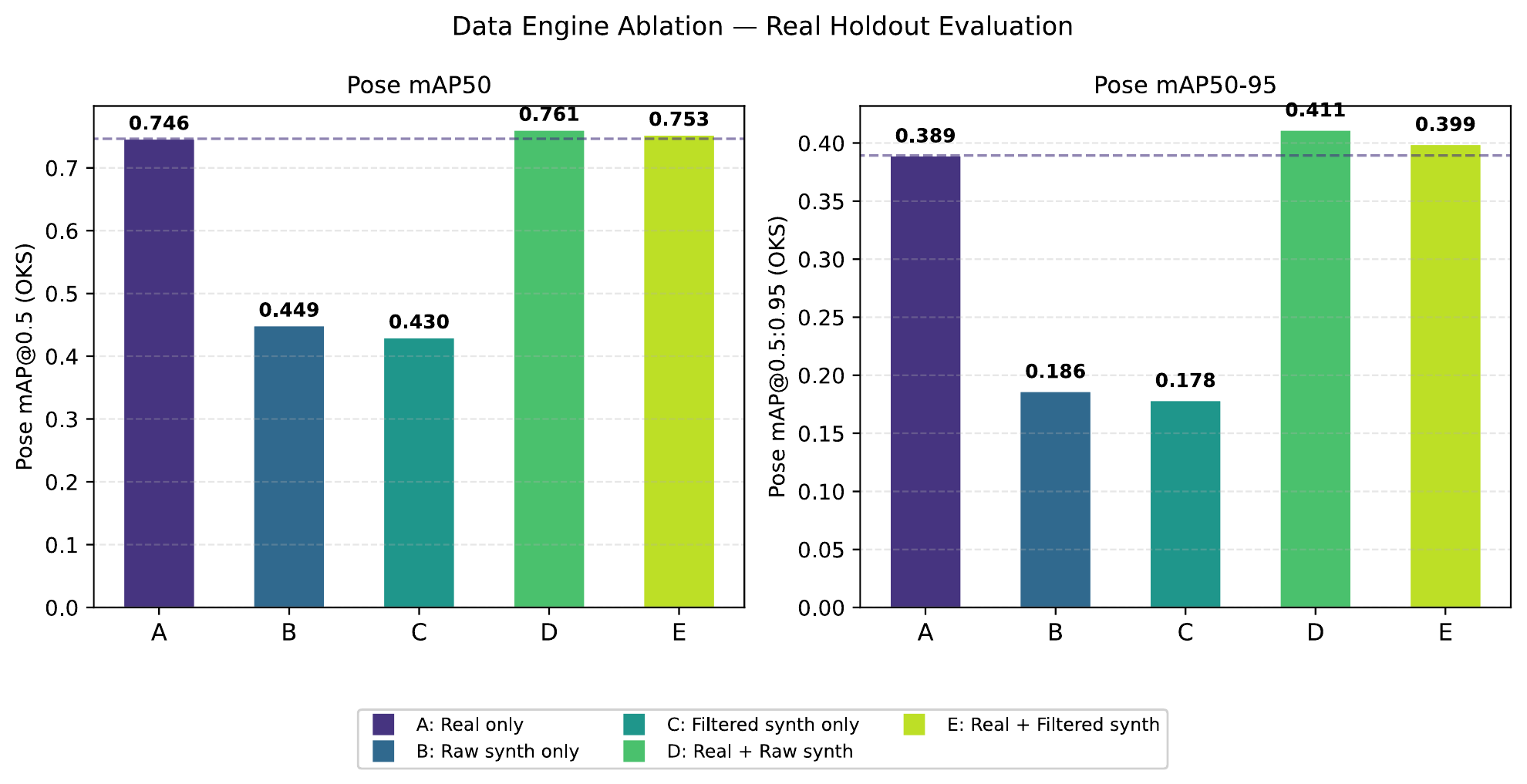}
  \caption{Pose $\mathrm{mAP}@0.5$ and $\mathrm{mAP}@0.5{:}0.95$ across the five training conditions on the shared real holdout set. Both mixed real+synthetic settings (D, E) score above the real-only baseline (A), while synthetic-only conditions (B, C) lag substantially behind. Each bar reflects a single training run; no seed variance is reported.}
  \label{fig:pose-ablation-dual}
\end{figure*}

To foreground the paper's main empirical message, we present the core pose comparison immediately after the evaluation protocol. Table~\ref{tab:pose-ablation} and Fig.~\ref{fig:pose-ablation-dual} summarize the central evidence that the mixed conditions achieve higher pose mAP than the real-only condition, while synthetic-only training still exhibits a clear real-domain gap.

\begin{table*}[!t]
  \centering
  \caption{Core pose comparison on the shared 280-image real holdout set. All models use identical YOLO11s-pose hyperparameters (100 epochs, 640-pixel inputs, batch size 16); data composition, total image count, and resulting update count differ by condition.}
  \label{tab:pose-ablation}
  \small
  \renewcommand{\arraystretch}{1.1}
  \setlength{\tabcolsep}{3.5pt}
  \begin{tabular}{@{}l l c c c c c c@{}}
    \toprule
     &  &  & \multicolumn{2}{c}{Pose mAP} & Box mAP &  &  \\
    \cmidrule(lr){4-5}\cmidrule(lr){6-6}
    Condition & Training data & Images & @0.5 & @0.5:0.95 & @0.5 & Precision & Recall \\
    \midrule
    A & Real only & 1589 & 0.746 & 0.389 & 0.868 & 0.800 & 0.724 \\
    B & Raw synth only & 479 & 0.449 & 0.186 & 0.624 & 0.666 & 0.399 \\
    C & Filtered synth only & 518 & 0.430 & 0.178 & 0.608 & 0.636 & 0.402 \\
    D & Real + raw synth & 2068 & \textbf{0.761} & \textbf{0.411} & \textbf{0.875} & \textbf{0.822} & 0.715 \\
    E & Real + filtered synth & 2107 & 0.753 & 0.399 & 0.866 & \textbf{0.822} & 0.713 \\
    \bottomrule
  \end{tabular}
\end{table*}

\begin{table}[t]
  \caption{Key implementation settings recorded in the released configuration files.}
  \label{tab:implementation-settings}
  \centering
  \small
  \renewcommand{\arraystretch}{1.15}
  \begin{tabular}{@{}p{0.28\linewidth} p{0.62\linewidth}@{}}
    \toprule
    Component & Setting \\
    \midrule
    Downstream learner & YOLO11s-pose; 100 epochs; 640-pixel inputs; batch size 16; patience 30. \\
    Semantic filter & SigLIP2 SO400M/16 NaFlex; semantic-margin threshold 0.9245. \\
    Structural filter & YOLO26x-pose; minimum person confidence 0.25; at least 12 keypoints above 0.5; bounding-box area ratio 0.05--1.00. \\
    Synthetic-label export & YOLO26x-pose; confidence 0.25; IoU 0.45; validation ratio 0.20; split seed 42. \\
    \bottomrule
  \end{tabular}
\end{table}

\subsection{Datasets and Setup}
\label{sec:exp-setup}

\paragraph{Real anchor and evaluation split.}
For the pose comparison, the real training set contains 1,589 COCO body-pose images~\cite{lin2014microsoft}, and all conditions are evaluated on the same 280-image real holdout set. This shared evaluation target prevents benchmark drift across conditions. Because training-set sizes and associated update counts differ, it does not by itself attribute performance differences to synthetic data.

\paragraph{Synthetic data generation and filtering.}
We generate pose-conditioned synthetic samples with a fixed pretrained diffusion model under structured control signals. We study two exported synthetic training pools produced in separate generation runs: an \textbf{unfiltered (raw)} variant with 479 images and a \textbf{curated} variant with 518 images. The curated pool is therefore not a subset of the raw pool, and their comparison is not a paired test of filtering.

\paragraph{Downstream model and training.}
We train YOLO11s-pose~\cite{ultralytics2024yolo11} under an identical hyperparameter recipe for every condition. Image resolution, nominal epoch count, batch size, augmentation policy, and evaluation procedure are kept fixed; data composition and dataset size differ, resulting in unequal optimizer updates. Table~\ref{tab:implementation-settings} records the key released settings.

\subsection{Compared Training Conditions}
\label{sec:exp-conditions}

Table~\ref{tab:pose-ablation} compares five conditions that characterize synthetic quantity, filtering configuration, and real--synthetic mixing:

\begin{itemize}
  \item \textbf{A: Real only} --- train on real COCO pose data only.
  \item \textbf{B: Raw synth only} --- train on unfiltered synthetic pose data only.
  \item \textbf{C: Filtered synth only} --- train on filtered synthetic pose data only.
  \item \textbf{D: Real + raw synth} --- augment the real set with raw synthetic images.
  \item \textbf{E: Real + filtered synth} --- augment the real set with filtered synthetic images.
\end{itemize}

Conditions B and C quantify the intrinsic domain gap of synthetic-only training, while D and E test whether synthetic data provides practical value as augmentation on top of real supervision. Synthetic keypoint labels require no additional manual annotation, but they are produced by the pose model described earlier and carry its errors, which we do not separately audit; we therefore avoid describing the pipeline as having negligible annotation cost.

\subsection{Core Pose Comparison Results}
\label{sec:exp-pose-ablation}

Table~\ref{tab:pose-ablation} summarizes the full five-condition comparison. The results support three observations, which we state with the strength a single-seed design allows.

\paragraph{Synthetic-only training underperforms real data by a large margin.}
Both synthetic-only conditions lag far behind the real-only baseline on the shared real holdout set. Raw synthetic only (B) reaches pose $\mathrm{mAP}@0.5=0.449$, and filtered synthetic only (C) reaches 0.430, compared with 0.746 for the real-only condition (A). This roughly 30-point deficit confirms a substantial domain gap: synthetic data alone cannot replace real human pose annotations in the current setting.

\paragraph{Adding synthetic data trends positive as annotation-efficient augmentation.}
When synthetic data is added to the real set, both mixed conditions achieve higher pose $\mathrm{mAP}@0.5$ and $\mathrm{mAP}@0.5{:}0.95$ than the real-only baseline. Condition D (real + raw synthetic) achieves the best overall result with pose $\mathrm{mAP}@0.5=0.761$ and pose $\mathrm{mAP}@0.5{:}0.95=0.411$, an absolute gain over A of $+0.015$ and $+0.022$ respectively, corresponding to relative increases of $2.1\%$ and $5.7\%$. We report the absolute differences alongside the relative ones because the relative figures overstate a change of this magnitude. This is the paper's central observation: \emph{under our protocol, synthetic augmentation trends positive for downstream pose estimation when anchored by real data}. The margin is small enough that we do not treat it as established; Section~\ref{sec:limitations} states what would be required to do so.

\paragraph{Filtering shows no measurable benefit at the current data scale.}
The curated exports do not outperform the unfiltered exports in this relatively small-scale comparison. In the synthetic-only setting, C is slightly below B, and in the mixed setting, E still improves over the real-only baseline but remains slightly below D (0.753 vs.\ 0.761 in pose $\mathrm{mAP}@0.5$). Because the curated pools are also \emph{larger} than the raw pools (518 vs.\ 479 in the synthetic-only setting, 2{,}107 vs.\ 2{,}068 in the mixed setting), this comparison confounds selection quality with pool size and is not a paired evaluation of the filter. With that caveat, the pattern suggests that the current curation signal is not yet strong enough to translate into a clear downstream advantage, even though the overall direction of the synthetic-augmentation effect remains positive.

\begin{figure}[t]
  \centering
  \includegraphics[width=\linewidth]{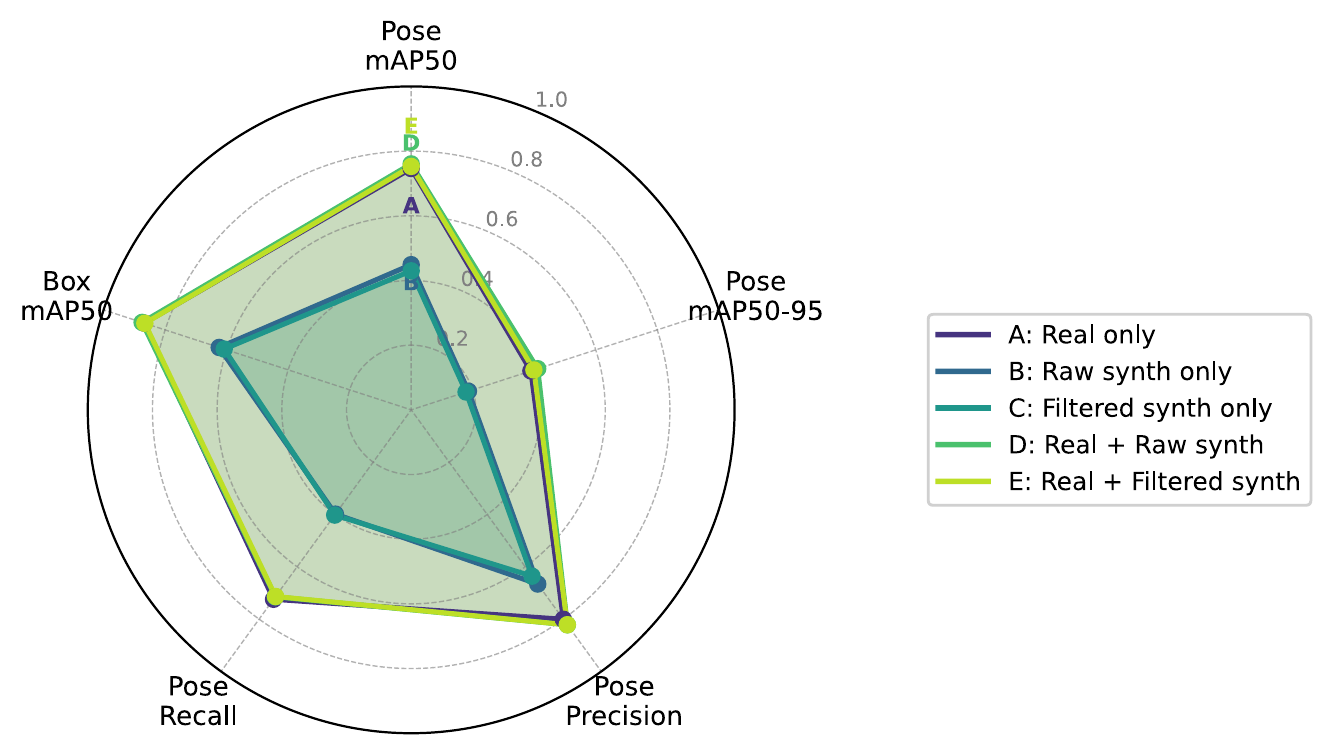}
  \caption{Five-metric radar profile for all training conditions. Mixed settings achieve the strongest pose mAP and precision values, while real-only training retains the highest recall. Synthetic-only conditions (B, C) remain weaker across all reported metrics, indicating that domain gap is the dominant limitation rather than any single metric artifact.}
  \label{fig:pose-ablation-radar}
\end{figure}

\paragraph{Discussion.}
Two aspects of the results warrant attention. First, the observed benefit is larger on the stricter metric: the best mixed setting gains $+0.015$ on pose $\mathrm{mAP}@0.5$ and $+0.022$ on pose $\mathrm{mAP}@0.5{:}0.95$ ($+2.1\%$ and $+5.7\%$ relative), which would be consistent with synthetic data contributing supervision beyond coarse keypoint localization --- though with one run per condition we cannot separate this pattern from noise. Second, conditions D and E use more total training images than A, so an unknown share of the difference is attributable to data volume rather than to the synthetic data itself. This confound is not resolved by appealing to the deployment scenario: even if additional labeled data is cheap to produce, the comparison as run does not isolate what the synthetic images contribute. We therefore treat the budget-matched comparison described in Section~\ref{sec:limitations} as required rather than optional, and we do not claim near-zero annotation cost, since synthetic keypoint labels inherit the errors of the pose model that produces them. Feature-space evidence complementing these findings is provided in Appendix A.

\section{Limitations}
\label{sec:limitations}

Four constraints bound the conclusions above, and we state them explicitly because each one is a prerequisite for a stronger claim rather than a peripheral caveat.

First, each condition is trained once. The $+0.015$ pose $\mathrm{mAP}@0.5$ difference between A and D is within the range we would expect from run-to-run variation in small-data fine-tuning, and we do not report seed variance. Establishing the effect requires three to five seeds per condition with means and confidence intervals.

Second, the mixed conditions see more images and more optimizer updates than the real-only baseline (2{,}068 and 2{,}107 images against 1{,}589). The observed difference therefore cannot be attributed to the synthetic data specifically rather than to additional training signal. A budget-matched control --- real-only oversampling at equal update counts, or replacement at a fixed total image count --- is needed to separate the two.

Third, our raw and curated pools differ in size (479 and 518 images), because they were exported from separate generation runs rather than filtered from one frozen parent pool. The raw-versus-curated comparison is consequently not a paired evaluation of the filter, and we draw no conclusion about curation quality from it. A proper test fixes a parent pool and compares random-$k$, semantic top-$k$, structural top-$k$, and two-stage top-$k$ at identical $k$.

Fourth, our structural filter and our downstream learner are both from the YOLO family, so retention may be biased toward samples this model family already handles well rather than toward generally informative ones. Verifying that the ranking transfers requires either an independent structural checker or a non-YOLO downstream pose model.

We also evaluate on a single 280-image holdout drawn from one dataset, which limits external validity and leaves open the possibility that threshold and ratio choices were tuned against the same split used for reporting.

\section{Conclusion}

We presented a threshold-calibrated synthetic-first data engine for low-data vision workflows and evaluated it with a five-condition pose comparison centered on real-domain generalization. The main empirical observation is that synthetic data is most useful when paired with real anchors, and least useful on its own. Both mixed settings achieve higher pose mAP than the real-only condition, while both synthetic-only settings remain substantially below real-only performance, confirming that domain gap is still the dominant limitation.

The results also refine the scope of the claims: filtering should be regarded as a scale-dependent design choice rather than a conclusively validated source of gain at the current data size, and the HITL feedback loop remains an implemented but unevaluated system extension.

Fig.~\ref{fig:pose-ablation-radar} contextualizes the pose-metric improvements with the tradeoffs in box mAP, precision, and recall; it should not be read as a common ordering across all metrics. Overall, the paper supports a conservative conclusion: under this single-run comparison, synthetic-first pipelines may offer annotation-efficient augmentation, but not a substitute for real data. Future work will focus on larger-scale synthetic generation, stronger data-budget controls, and more targeted validation of when curation and human feedback materially improve real-domain performance.

\appendix

\section{Feature-Space Domain Gap Visualization}
\label{sec:appendix-tsne}

\begin{figure}[t]
  \centering
  \includegraphics[width=\linewidth]{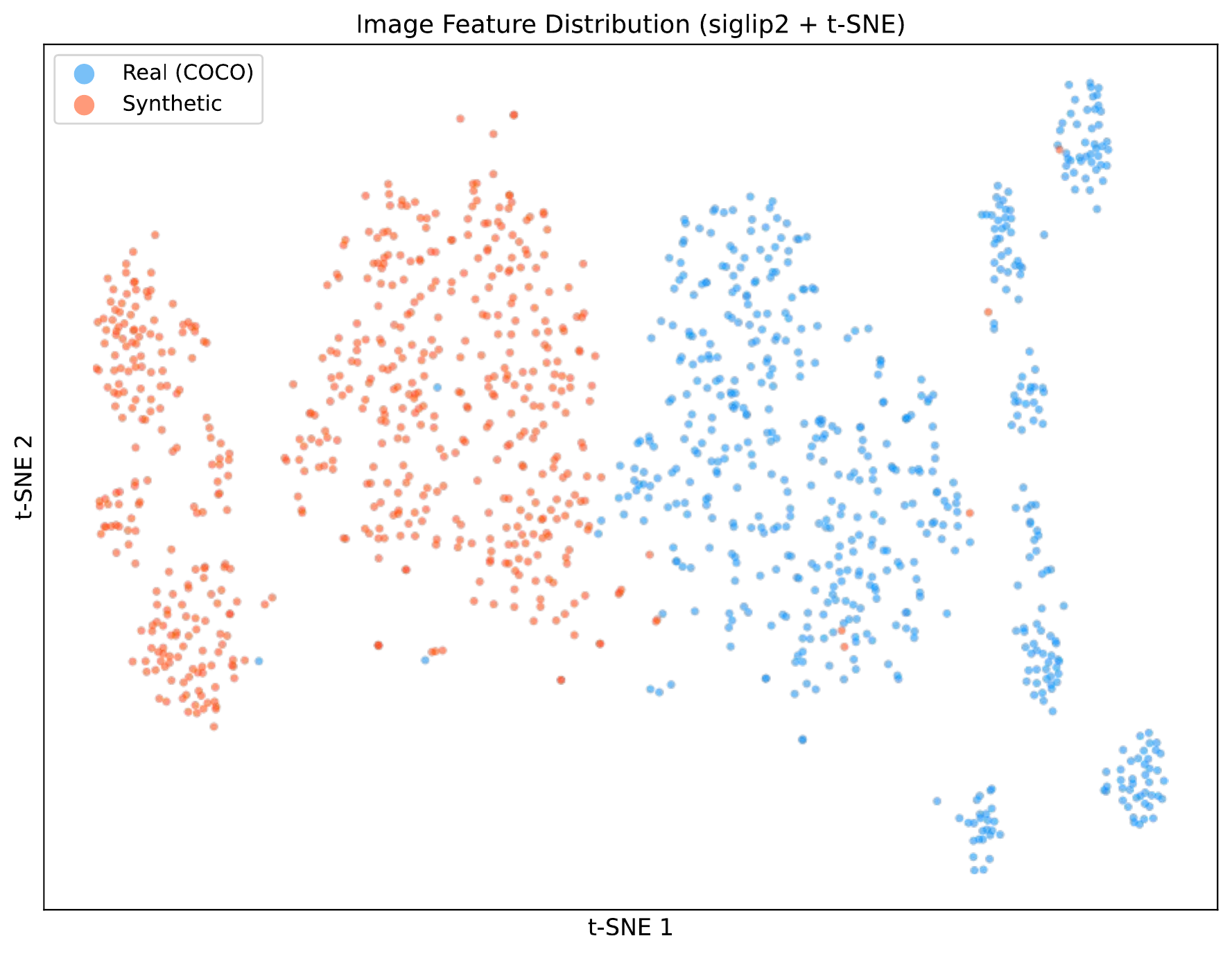}
  \caption{Feature-space visualization of real and synthetic samples for the pose task. The remaining separation between domains is consistent with the strong gap observed under synthetic-only training, while the mixed-data gains suggest that useful complementary coverage still exists.}
  \label{fig:pose-tsne}
\end{figure}

Figure~\ref{fig:pose-tsne} provides a descriptive feature-space view of the domain gap between real and synthetic samples for the pose task. The SigLIP2-embedding t-SNE projection shows incomplete overlap, consistent with the weak absolute performance of conditions B and C on the real holdout set. Because t-SNE depends on the embedding and projection settings, we treat this figure as a qualitative diagnostic rather than an independent distributional test. The mixed-data pose-metric differences suggest that full distribution matching may not be necessary for synthetic data to be useful when grounded by real anchors.

\bmhead{Acknowledgments}
This work was supported in part by the U.S. National Science Foundation under Grants No. 2146497, 2244219, 2315595, 2315596, and 2416872. Any opinions, findings, and conclusions or recommendations expressed in this material are those of the author and do not necessarily reflect the views of the National Science Foundation.

\bmhead{Declarations}

\bmhead{Competing Interests}
On behalf of all authors, the corresponding author states that there is no conflict of interest.

\bmhead{Funding Information}
This work was supported in part by the U.S. National Science Foundation under Grants No. 2146497, 2244219, 2315595, 2315596, and 2416872.

\bmhead{Author Contribution}
All authors contributed to the study conception and design, interpretation of the results, and revision of the manuscript. All authors read and approved the final manuscript.

\bmhead{Data Availability Statement}
The COCO dataset used in this study is publicly available. The implementation and configuration files for the reported experiments are publicly available at \url{https://github.com/Yan-s-Lab/Data-Engine}.

\bmhead{Research Involving Human and/or Animals}
Not Applicable. This study uses publicly available datasets and does not involve human participants or animals as research subjects.

\bmhead{Informed Consent}
Not Applicable.

\bibliography{references}

\end{document}